\documentstyle[12pt]{article}

\newlength{\myleftmargin}
\newlength{\paperwidth}
\begin{document}
\renewcommand{\thefootnote}{\fnsymbol{footnote}}
\begin{flushright}
KOBE--FHD--94--01\\
February~~24~~~1994
\end{flushright}
\begin{center}
{\large \bf A Way to Measure Polarized Gluon
Distributions\footnote[2]{Talk presented at meeting on ``Particle
Physics and its Future" (Yukawa Inst. for Theor. Phys., Kyoto,
Jan. 17 -- 21, 1994).}}\\

\vspace{3em}

T. Morii$^{1, 2}$, S. Tanaka$^1$ and T. Yamanishi$^2$\\

\vspace{1em}

Faculty of Human Development, Division of Sciences for Natural Environment,\\
Kobe University, Nada, Kobe 657, Japan$^1$\\
Graduate School of Science and Technology,
Kobe University, Nada, Kobe 657, Japan$^2$\\

\vspace{4em}

{\bf Abstract}
\end{center}

Effects of the spin--dependent gluon distributions on J/$\psi$ productions in
polarized ep and pp collisions are investigated. These productions serve as a
very clean probe of the spin--dependent gluon distributions in a proton.

\baselineskip=20pt

\vspace{5em}

There have been several theoretical approaches to get rid of {\it so--called}
the ``proton spin
crisis''\cite{crisis,Stirling}. Among them, there is an interesting idea that
gluons contribute significantly to the proton spin through the U${}_A(1)$
anomaly\cite{Altarelli}. In this description the first moment of the
spin--dependent gluon distribution ($\Delta G(Q^2)$) inside a proton is as
large as $5\sim 6$ at $Q^2_0=10.7$GeV$^2$(EMC value) and concomitantly the
amount of the proton spin carried by quarks does not necessarily become small.
However, is the polarized gluon contribution really so large in a proton?
In order to confirm this, it is absolutely necessary to measure, in
experiments, the physical quantities which are sensitive to the magnitude
of spin--dependent gluon distribution.

In the present work, we study two interesting processes which
predominantly depend on the spin--dependent gluon distributions : one is the
J/$\psi$ production in polarized proton--polarized
proton collisions and the other is the inelastic J/$\psi$ production in
polarized electron--polarized proton collisions \cite{Morii}.
In order to discuss how these processes are affected by
spin--dependent gluon distributions, we take the following three different
types of $x\delta G(x)$ :
\begin{enumerate}
\renewcommand{\labelenumi}{(\alph{enumi})}
\item our model \cite{Morii1}~;
\begin{eqnarray}
&&x\delta G(x, Q^2=10.7{\rm GeV}^2)=3.1x^{0.1}(1-x)^{17}
{}~{\rm then}~~~~~~~~~~~~
\label{eqn:typeA}\\
&&\Delta G(Q^2_{EMC})=6.32~.\nonumber
\end{eqnarray}
\item Cheng--Lai type model \cite{HYCheng}~;
\begin{eqnarray}
&&x\delta G(x, Q^2=10{\rm GeV}^2)=3.34x^{0.31}(1-x)^{5.06}(1-0.177x)
{}~{\rm then}
\label{eqn:typeB}\\
&&\Delta G(Q^2_{EMC})=5.64~.\nonumber
\end{eqnarray}
\item no gluon polarization model \cite{HYCheng}~;
\begin{equation}
x\delta G(x, Q^2=10{\rm GeV}^2)=0~~{\rm then}~~\Delta G(Q^2_{EMC})=0~.
\label{eqn:typeC}
\end{equation}
\end{enumerate}
The behavior of $x\delta G(x, Q^2)$ and $\delta G(x, Q^2)/G(x, Q^2)$ are
depicted in Fig.1 (A) and (B) as a function of $x$, respectively.
As shown here, the $x\delta G(x)$ of type (a) has a sharp peak at $x<0.01$
and rapidly decreases with increasing $x$ while that of (b) has a peak at
$x\approx 0.05$ and gradually decreases with $x$.
The shape of $x\delta G(x)$ taken by
many authors\cite{gluons} is almost the same as that of type (b).

First, we discuss the inclusive J/$\psi$
production in polarized proton--polarized proton collisions. Since the
J/$\psi$ productions come out only via gluon--gluon fusion processes at the
lowest order of QCD diagrams, those cross sections are sensitive to the
spin--dependent gluon distribution in a proton. Let us define the two--spin
asymmetry $A^{J/\psi}_{LL}(pp)$ for this process by
\begin{equation}
A^{J/\psi}_{LL}(pp)=\frac{\left[d\sigma(p_+ p_+\rightarrow J/\psi ~X)
-d\sigma(p_+ p_-\rightarrow J/\psi ~X)\right]}
{\left[d\sigma(p_+ p_+\rightarrow J/\psi ~X)+
d\sigma(p_+ p_-\rightarrow J/\psi ~X)\right]}
=\frac{Ed\Delta\sigma/d^3p}{Ed\sigma/d^3p}~,
\label{eqn:ALL}
\end{equation}
where
$p_+ (p_-)$ denotes that the helicity of a proton is positive (negative).
In eq.(\ref{eqn:ALL}), the numerator (denominator) represents the
spin--dependent (spin--independent) differential cross section for the
hard--scattering parton model and is formulated in the
framework of perturbative QCD\cite{Gastmans}.
For estimation of $A_{LL}^{J/\psi}(pp)$, we take the
spin--dependent gluon distributions (a), (b) and (c) given by
eqs.(\ref{eqn:typeA}), (\ref{eqn:typeB}) and (\ref{eqn:typeC}).
Setting $\theta=90^{\circ}$ ($\theta$ is the production angle
of J/$\psi$ in the CMS of colliding protons)
and using the spin--independent gluon
distribution function of the DO parametrization \cite{Duke} for (a),
and the DFLM parametrization \cite{DFLM} for (b) and (c), we have
calculated $A_{LL}^{J/\psi}(pp)$ for several choices of $Q^2$ ;
$Q^2=m^2_{J/\psi}+p_T^2$, $4p_T^2$, $(\hat s\hat t\hat u)^{1/3}$, $-\hat t$
and so on.
We see that $A_{LL}^{J/\psi}(pp)$ for each type of the spin--dependent
gluon distributions is insensitive to the choice of $Q^2$.
Thus, we here take $Q^2=m_{J/\psi}^2+p_T^2$ by taking the mass effect of the
J/$\psi$ particle into account. The results of $A_{LL}^{J/\psi}(pp)$ are
shown in Fig.2 as a function of $p_T$ of the J/$\psi$ at (A) $\sqrt s=20$ and
(B) $100$ GeV. At $\sqrt s=20$ GeV our largely polarized gluon distribution,
(a), contributes little to
$A_{LL}^{J/\psi}(pp)$ in all $p_T$ regions because the
region near the peak of $x\delta G(x)$ is kinematically cut.
The $A_{LL}^{J/\psi}$ predicted with type (a) is not so significantly
different from that with no gluon polarization (type (c)), and it is
practically difficult to find the difference between them.
However, for higher energies such as $\sqrt s=100$ GeV, we might distinguish
types (a) from (c) for spin--dependent gluon distributions by choosing a
moderate $p_T$ region.
In addition, one can see that the behavior of $A_{LL}^{J/\psi}$ for type (b)
largely differs from those for types (a) and (c) at $\sqrt s=20$ and $100$ GeV.
Therefore, it is expected that one can either rule out or confirm type (b) by
measuring $A_{LL}^{J/\psi}$.

Next, in order to distinguish types (a) from (c) of $x\delta G$, we consider
inelastic J/$\psi$ productions in polarized ep collisions \cite{Morii}.
In the inelastic region where the J/$\psi$ particles are produced via the
photon--gluon fusion, $\gamma^* g \rightarrow J/\psi ~g$,
the spin--dependent differential cross section is given by
\begin{equation}
\frac{d\Delta\sigma}{dx} = x\delta G(x, Q^2) \delta f(x, x_{min})~,
\label{eqn:ddsdx}
\end{equation}
where $\delta G(x, Q^2)$ is the spin--dependent gluon distribution
function and $x$ the fraction of the proton momentum carried by the initial
state gluon.  $\delta f$ is a function which is
sharply peaked at $x$ just above $x_{min}$ and given by \cite{Morii}
\begin{eqnarray}
\delta f(x, x_{min})&=&\frac{16\pi\alpha_S^2\Gamma_{ee}}{3\alpha m_{J/\psi}^3}
\frac{x_{min}^2}{x^2} \label{eqn:df}\\
&\times&\left[\frac{x-x_{min}}{(x+x_{min})^2}+
\frac{2x_{min}x\ln\frac{x}{x_{min}}}{(x+x_{min})^3}-
\frac{x+x_{min}}{x(x-x_{min})}+
\frac{2x_{min}\ln\frac{x}{x_{min}}}{(x-x_{min})^2}\right]~,\nonumber
\end{eqnarray}
where $x_{min}\equiv m_{J/\psi}^2/s_T$ and $\sqrt {s_T}$ is the total energy
in photon--proton collisions.
Fig.3 shows the $x$ dependence of $d\Delta\sigma/dx$ calculated with types
of (a) and (b) for various energies including relevant HERA energies.
As $\delta f$ has a sharp peak, the observed cross section
$d\Delta\sigma/dx$ directly reflects the spin--dependent gluon distribution
near $x_{peak}$. As is seen from eq.(\ref{eqn:ddsdx}), $d\Delta\sigma/dx$ is
linearly dependent on the spin--dependent gluon distribution. Thus, if
$\delta G(x)$ is small or vanishing, $d\Delta\sigma/dx$ must be necessarily
small.
We are eager for the result given in Fig.3 being checked in the forthcoming
experiments.

In summary, we have examined the effect of the gluon polarization on
some physical quantities in special processes which are sensitive to the
polarized gluon distribution.
As for $A_{LL}^{J/\psi}(pp)$, there would be a good chance to find an evidence
of largely polarized gluons in moderate $p_T$ regions ($p_T > 1$GeV).
Furthermore, since $d\Delta\sigma/dx$ for J/$\psi$ leptoproductions is directly
proportional to
the polarized gluon distribution, one can easily examine the magnitude of
the gluon polarization by measuring these quantities in experiments.
The J/$\psi$ productions in polarized ep and pp collisions
considered here can therefore serve as a very clean probe of the polarized
gluon distributions in a proton.

\vfill\eject

\vfill\eject

\begin{center}
{\large \bf Figure captions}
\end{center}
\begin{description}
\item[Fig. 1:] The $x$ dependence of (A) $x\delta G(x, Q^2)$ and (B)
$\delta G(x, Q^2)/G(x, Q^2)$ for various types (a)--(c) given by
eqs.(\ref{eqn:typeA}), (\ref{eqn:typeB}) and (\ref{eqn:typeC}) at
$Q^2=10.7$ GeV$^2$.

\vspace{2em}

\item[Fig. 2:] The two--spin asymmetries $A_{LL}^{J/\psi}(pp)$ for
$\theta=90^{\circ}$ calculated by
using types (a), (b) and (c) as the spin--dependent gluon distribution
functions, as a function of transverse momenta $p_T$ of J/$\psi$
at (A) $\sqrt s=20$ GeV, and (B) $\sqrt s=100$ GeV.
The solid, dashed and dash--dotted curves correspond to types (a), (b) and (c),
respectively. $Q^2$ is typically taken to be $m^2_{J/\psi}+p_T^2$.

\vspace{2em}

\item[Fig. 3:] The distribution $d\Delta\sigma/dx$ predicted by using types
(a) and (b) of $x\delta G(x, Q^2)$, as a function of $x$ for different values
of $\sqrt{s_T}$. The solid (dashed) curve corresponds to type (a) ((b)).
\end{description}

\begin{thebibliography}{1}
\bibitem{Morii1}
This is a modified version of $x\delta G(x, Q^2)$ proposed in Ref.\cite{Morii}.
\end{thebibliography}
\end{document}